\documentclass[aps,pra,10pt,notitlepage,nofootinbib,tightenlines,floatfix,
twocolumn,superscriptaddress]{revtex4-2}
\UseRawInputEncoding

\usepackage[english]{babel}
\usepackage[margin=50pt]{geometry}

\usepackage[colorlinks=true, allcolors=blue]{hyperref}
\hypersetup{
    colorlinks=true,       % false: boxed links; true: colored links
    linkcolor=red,          % color of internal links
    citecolor=magenta,        % color of links to bibliography
    filecolor=magenta,      % color of file links
    urlcolor=cyan,           % color of external links
    runcolor=cyan
}
\usepackage{bm,bbm,amssymb,amsmath,amsfonts,amsthm,mathrsfs,MnSymbol,times}
\usepackage{algorithm2e}
\usepackage[normalem]{ulem}

\usepackage{bibunits}
\usepackage{bbm}

\usepackage{graphicx}
\usepackage{caption}
\usepackage{subcaption}
\usepackage{amsmath}
\usepackage{color}
\usepackage{xcolor}
\usepackage{bbold}
\usepackage[utf8]{inputenc}
\usepackage{empheq}
\usepackage{braket}
\usepackage{enumerate}

\newcommand{\seq}[1]{\begin{align}\begin{split}#1\end{split}\end{align}}

\newcommand{\bel}{\begin{easylist}[itemize]}
\newcommand{\eel}{\end{easylist}}

\newcommand{\sz}{\sigma_z}

\newcommand{\mrm}{\mathrm}

\newcommand{\tr}{\mathrm{Tr}}
\newcommand{\re}{\mathrm{Re}}
\newcommand{\im}{\mathrm{Im}}

\newcommand{\SNR}{\mathrm{SNR}}

\renewcommand{\l}{\left}
\renewcommand{\r}{\right}
\graphicspath{ {figures/} }

\begin{document}

\title{Fast charge noise sensing using a spectator valley state in a singlet-triplet qubit}

\author{David W.~Kanaar$^*$}
\affiliation{Department of Electrical and Computer Engineering,
University of California, Los Angeles, Los Angeles, CA 90095, USA}
\affiliation{Center for Quantum Science and Engineering, University of California, Los Angeles, Los Angeles, CA 90095, USA}
\author{Yasuo Oda$^*$}
\affiliation{Department of Physics, University of Maryland Baltimore County, Baltimore, MD 21250, USA}
\author{Mark F.~Gyure}
\affiliation{Department of Electrical and Computer Engineering,
University of California, Los Angeles, Los Angeles, CA 90095, USA}
\affiliation{Center for Quantum Science and Engineering, University of California, Los Angeles, Los Angeles, CA 90095, USA}
\author{J.~P.~Kestner}
\affiliation{Department of Physics, University of Maryland Baltimore County, Baltimore, MD 21250, USA}

\begin{abstract}
    Semiconductor spin qubits are a promising platform for quantum computing but remain vulnerable to charge noise. Accurate, in situ measurement of charge noise could enable closed-loop control and improve qubit performance. Here, we propose a method for real-time detection of charge noise using a silicon singlet-triplet qubit with one electron initialized in an excited valley state. This valley excitation acts as a spectator degree of freedom, coupled to a high-quality resonator via the exchange interaction, which is sensitive to charge-noise-induced voltage fluctuations. Dispersive readout of the resonator enables a continuous, classical measurement of exchange fluctuations during qubit operation. Signal-to-noise analysis shows that, under realistic device parameters, sub-millisecond measurement times are possible using a quantum-limited amplifier. Even without such an amplifier, similar performance is achievable with appropriately engineered resonator parameters. This approach allows the probe to monitor slow drift in exchange in real time, opening the door to feedback and feedforward strategies for maintaining high-fidelity quantum operations. Importantly, the protocol preserves spin coherence and can be run concurrently with qubit logic gates.
\end{abstract}

\maketitle
\def\thefootnote{*}\footnotetext{These authors contributed equally to this work.}

\section{Introduction}

Semiconductor spin qubits are a promising platform for quantum computing because of their compatibility with industrial CMOS fabrication techniques~\cite{zwanenburg_silicon_2013,maurand_cmos_2016,gonzalez-zalba_scaling_2021}. In recent years, impressive progress has been made, with gate fidelities reaching  the thresholds needed for quantum error correction in few qubit devices. 
However, achieving full fault-tolerant operation will require not only more qubits, but also further suppression and stabilization of environmental noise sources~\cite{tanttu_assessment_2024}. \par
In fact, a key challenge in silicon spin qubits is charge noise, which induces fluctuations in both the exchange interaction between spins~\cite{culcer_dephasing_2009,keith_impact_2022} and the Larmor frequency via spin-orbit coupling~\cite{ruskov_electron_2018}. These fluctuations limit gate fidelity and coherence times~\cite{tanttu_assessment_2024}, and by mitigating them at the physical qubit level reduces the overhead of error correction. The power spectral density of charge noise in these devices is typically measured through transport via a nearby sensor dot or quantum point contact~\cite{freeman_comparison_2016,yoneda_quantum-dot_2018,connors_low-frequency_2019,struck_low-frequency_2020,connors_charge-noise_2022,esposti_low_2023}. These measurements, however, do not occur during qubit operation and rely on spatial noise correlations which, though they may be strong, are far from perfect~\cite{boter_spatial_2020,ye_characterization_2024,rojas-arias_inferring_2025,yoneda_noise-correlation_2023}. Other noise measurements or calibration techniques, such as gate set tomography or real-time Hamiltonian estimation~\cite{shulman_suppressing_2014,huang_fidelity_2019,nakajima_coherence_2020,kim_approaching_2022,vepsalainen_improving_2022}, involve projective measurements on the qubit state which can be slow and are not compatible with continuous qubit operation. \par

In this work, we build on the idea of Ref.~\cite{kanaar_proposed_2024} and propose a new method for continuous, real-time measurement of charge noise using the valley degree of freedom of a silicon singlet-triplet (ST) spin qubit, with one electron intentionally loaded into an excited valley state. 

In the strained silicon quantum wells used for qubits, the valley states arise from the non-broken degeneracy of the conduction band minima. This degeneracy is lifted by valley-orbit coupling, resulting in the two valley eigenstates separated by an energy known as the valley splitting. The two electrons, one in an excited valley state, form a two-level system with an energy difference of approximately the difference between their valley splittings. The coupling between these energy levels depends on the exchange coupling, which fluctuates due to charge noise. 
By dispersively coupling this system to a resonator, a classical reflectometry measurement of the resonator can be performed with negligible effect on the spin qubit state. \par

Essentially, the valley degree of freedom acts as a natural ``spectator qubit" living on the same pair of electrons as the ST qubit itself. This colocated sensor is ideal from a resource standpoint and ensures that the measured noise is representative of what the spin qubit experiences. Because valley relaxation times in silicon can exceed 10~ms~\cite{penthorn_direct_2020}, orders of magnitude longer than typical gate durations, the valley state remains coherent and sensitive to low-frequency charge noise on timescales relevant for control. 
In contrast to a previous proposal using a similar idea for a single-electron (Loss-DiVincenzo) spin qubit that required Ge doping in the Si quantum well~\cite{kanaar_proposed_2024}, for the ST qubit we show here that standard existing Si heterostructures suffice and one no longer needs an electric dipole coupling between valley states.
This approach enables direct detection of fluctuations in the exchange and can be used to actively cancel out slow drifts in the exchange, enabling a form of closed-loop control that could improve qubit coherence and gate fidelity during extended operations.
 \par

The outline of the paper is as follows. In Sec.~\ref{sec:Model}, we derive the Hamiltonian for the singly-excited valley subspace coupled to a resonator. Using this Hamiltonian, in Sec.~\ref{sec:protocol} we compute the signal-to-noise ratio to determine the achievable measurement speed. In Sec.~\ref{sec:probeQubitOperation} we evaluate the feasibility of operating the qubit and the probe simultaneously and outline a protocol for feedback control. Finally, we discuss the results and conclude the manuscript in Sec.~\ref{sec:Summary}.

\section{Model}
\label{sec:Model}
\begin{figure*}[t!]
    \centering
    \begin{subfigure}{0.48\textwidth}
        \centering
        \includegraphics[width=\linewidth]{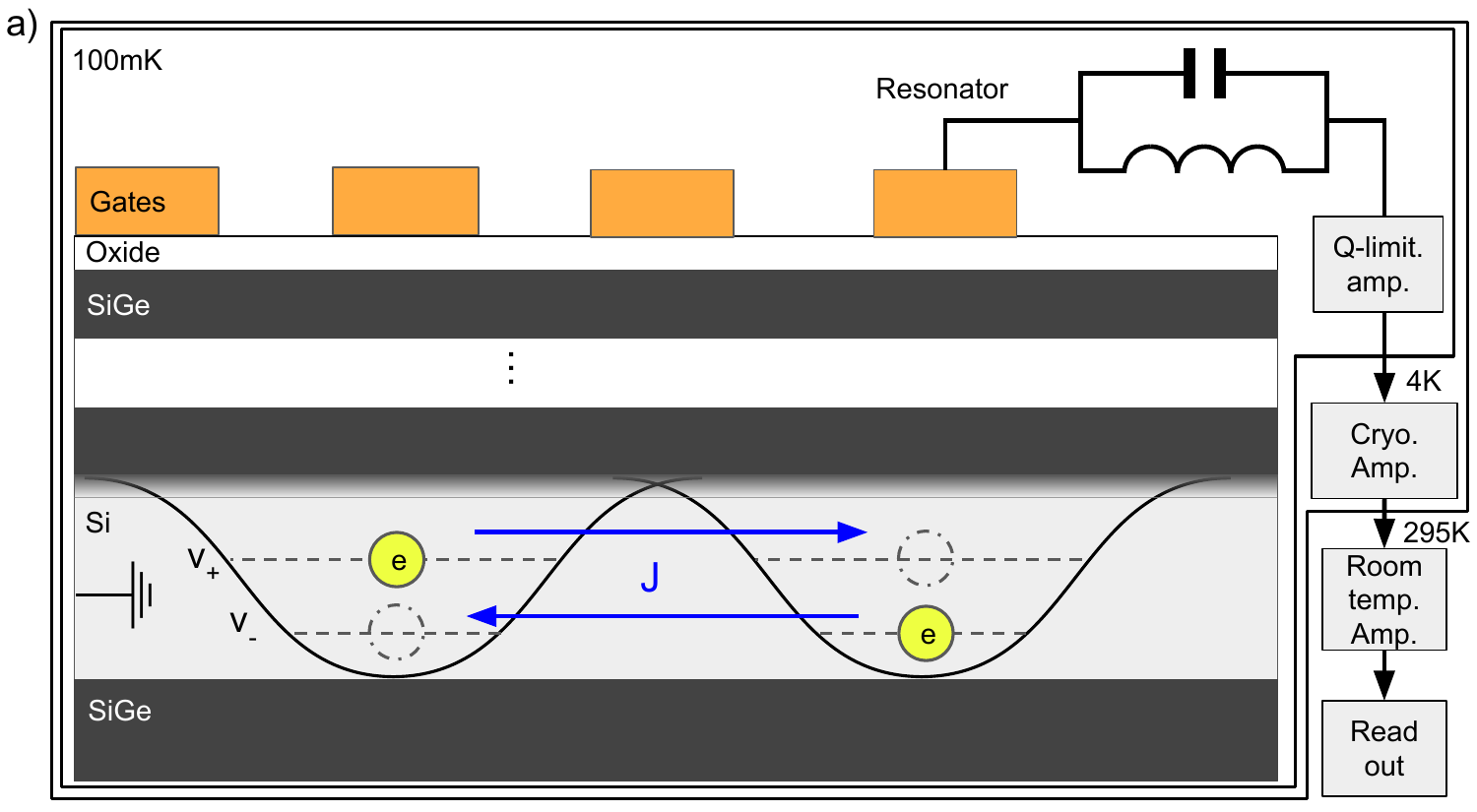} 
        \label{fig:subDevice1} 
    \end{subfigure}
    \hfill
    \begin{subfigure}{0.48\textwidth} 
        \centering
        \includegraphics[width=\linewidth]{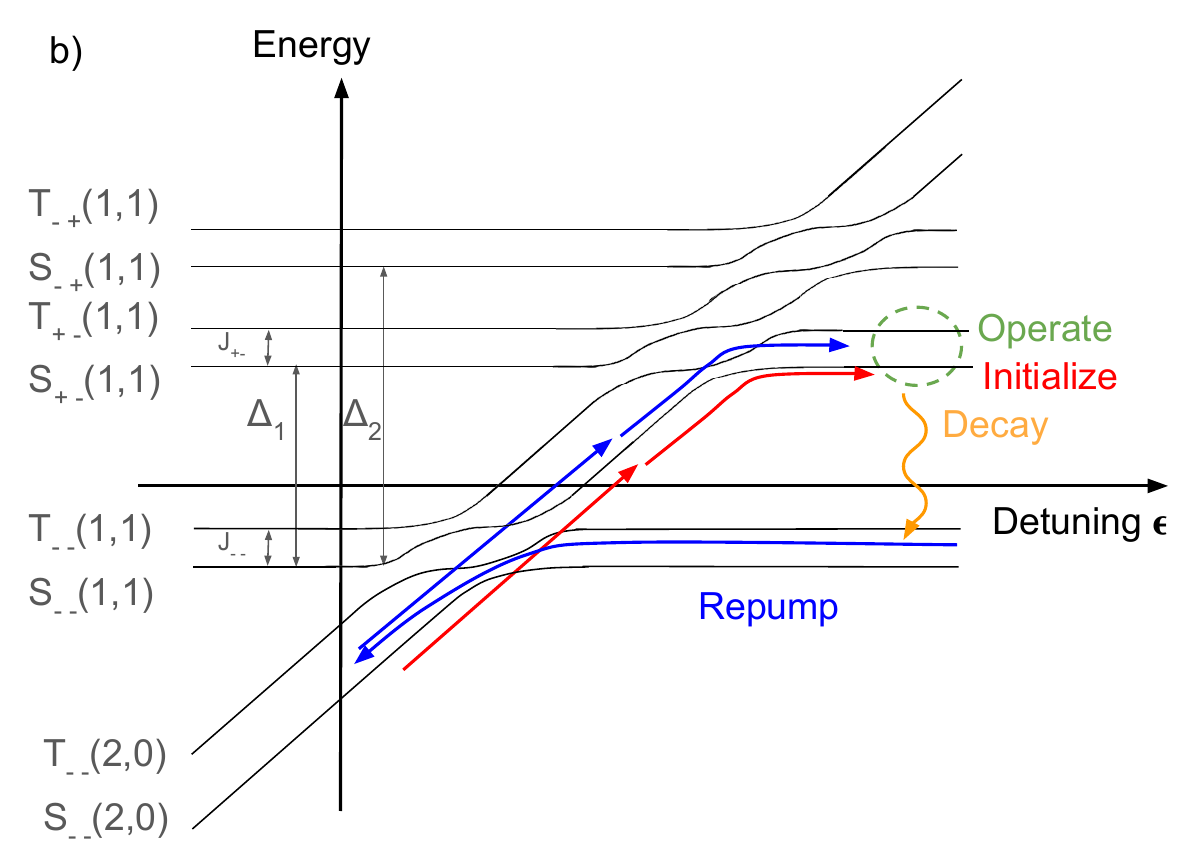} 
        \label{fig:subDevice2}
    \end{subfigure}
    \captionsetup{justification=centerlast,singlelinecheck=false}
    \caption{a) Schematic view of the proposed probe with two electrons, $e$, in two quantum dots with one electron in the excited valley state. Readout is done through a resonator and amplification chain shown on the side. $v_{\pm}$ indicate the valley states which are coupled by the exchange interaction $J$. b) Energy level diagram as a function of detuning, $\epsilon$, indicating the method for initialization and repumping after a valley decay. S$_{+ -}$ and T$_{+ -}$ indicate the singlet an triplet states with a + and - indicating an excited or ground valley. (i,j) inidicates i electrons in the left dot and j electrons in the right dot. For simplicity, polarized triplets are not shown.} 
    \label{fig:device}
\end{figure*}
Figure \ref{fig:device} a) illustrates the proposed setup for probing charge noise, consisting of a resonator coupled to two quantum dots, each containing one electron. The probing mechanism relies on initializing one of the electrons in an excited valley state. The energy level diagram depicting the initialization sequence is shown in Fig.~\ref{fig:device} b). The process begins by loading two electrons in a single dot in the singlet spin state. The detuning, $\epsilon$, is then diabatically ramped through the avoided crossing of the initial (2,0) singlet with the (1,1) singlet having no valley excitations and up to near the avoided crossing with the (1,1) singlet having one excited valley. Following this, the detuning is adiabatically ramped through that avoided crossing into the singly-excited-valley ST operating subspace. In the event of a valley decay of the ST into the unexcited-valley subspace, the system can be repumped into the operating subspace via a similar path, beginning with an adiabatic transition to the (2,0) ST subspace. While the numerical values in this manuscript are based on a Si/SiGe heterostructure, the concept is equally applicable to Si-MOS devices.\par
In this section, we derive the Hamiltonian for the system required to describe the probe. First, the Hamiltonian for the singly excited valley subspace is derived in Sec.~\ref{subsec:valleyH}, and conveniently rewritten in terms of a spin-dependent hybridized valley state in Sec.~\ref{subsec:HV}. This is then extended in Sec.~\ref{subsec:resonatorH} to include the resonator coupling to the valley states, resulting in the total probe Hamiltonian. Lastly, in Sec.~\ref{subsec:res_dyn_hom_meas} we discuss the resonator dynamics in the dispersive regime under homodyne measurement. In the equations below we use units such that $\hbar=1$.\par

\subsection{Valley subspace Hamiltonian derivation}\label{subsec:valleyH}
To derive the Hamiltonian of this system we start with a Fermi-Hubbard model that includes doubly occupied dots, as well as spin and valley states, following an approach similar to that in Ref.~\cite{tariq_impact_2022}. For simplicity, we assume that the orbital energies, on the order of $\sim$ meV, are sufficiently high to exclude contributions of the excited orbital states. The Fermi-Hubbard Hamiltonian is
\begin{equation}\label{eq:H0}
    \begin{split}
              H_0=&\sum_{v,s} t_c \left(c_{1,s,v}^{\dagger}c_{2,s,v} + c_{2,s,v}^{\dagger}c_{1,s,v}\right) \\
              &+ \sum_{i,v} \left(\frac{E_z^{(i)}}{2} \sigma_{z}^{(i)} + U_i n_{i,\uparrow} n_{i,\downarrow}+ \sum_{s=\uparrow,\downarrow} \mu_i n_{i,s}\right) \\ 
       & +\sum_{i=1}^2 \sum_{s=\uparrow,\downarrow}\Delta_i \left(e^{i \phi_i} c_{i,s,1}^{\dagger}c_{i,s,2}+e^{-i \phi_i} c_{i,s,2}^{\dagger}c_{i,s,1}\right),
    \end{split}
\end{equation}
where $t_c$ is the tunneling coupling between the dots, $c_{i,s,v}$ is the electron annihilation operator for the $i$-th dot with spin $s$ and valley $v$, $E_z^{(i)}$ is the Zeeman energy of an electron in the $i$-th dot, $\sigma_{z}^{(i)}$ is the Pauli $Z$ operator for the spin in the $i$-th dot, $\mu_i$ is the on-site energy of a single electron in the $i$-th dot, $U_i$ is the onsite Coulomb energy of the doubly occupied $i$-th dot, and $\Delta_i$ ($\phi_i$) is the valley coupling amplitude (phase) for an electron in the $i$-th dot. \par
Next, $H_0$ is transformed to the eigenbasis of the $\pm z$ valleys, which diagonalize the last line of Eq.~\eqref{eq:H0}. This results in 
\begin{equation}\label{eq:H0v}
    \begin{split}
          H_{0,\overline{v}}=&\sum_{\overline{v}=1,-1} \left(t_{+} c_{1,s,\overline{v}}^{\dagger}c_{2,s,\overline{v}} +t_{-} c_{1,s,\overline{v}}^{\dagger}c_{2,s,-\overline{v}} + \text{H.c.}\right) \\
          &+\hspace{-1pt} \sum_{i,\overline{v}} \hspace{0pt} \left(\frac{E_z^{(i)}}{2} \sigma_{z}^{(i)} + U_i n_{i,\uparrow} n_{i,\downarrow}+\Delta_i \tau_{z}^{(i)} + \hspace{-3pt} \sum_{s=\uparrow,\downarrow}  \hspace{-3pt}\mu_i n_{i,s}\right), \\ 
    \end{split}
\end{equation}
where $\overline{v}$ are the valley eigenstates, $t_{\pm}=\frac{t_c}{2}(1\pm e^{i \delta \phi })$ with $\delta \phi=\phi_1-\phi_2$, H.c.~is the Hermitian conjugate, and $\tau_{z}^{(i)}$ is the Pauli $Z$ operator for the valley eigenstates \cite{tariq_impact_2022}. \par 
$H_{0,\overline{v}}$ is then simplified by performing a Schrieffer-Wolff transformation, which removes the high-energy doubly occupied states. Additionally, the Hamiltonian is also moved to the singlet-triplet (ST) spin basis. Applying these transformations yields
\begin{equation}\label{eq:HSW1}
    \begin{split}
        H_{\text{ST},\overline{v}}=& \frac{\delta E_z}{2} \sigma_{x} + \Delta_i \tau_{z}^{(i)} \\
        &+\frac{J}{4} \sigma_{z} \left[  1+\tau_{z}^{(1)} \tau_{x}^{(2)}+ \sin(\delta \phi) \l(\tau_{y}^{(1)} \tau_{z}^{(2)}-\tau_{z}^{(1)} \tau_{y}^{(2)}\r) \right. \\
        &\left. +\cos(\delta \phi) (\tau_{y}^{(1)} \tau_{y}^{(2)}+\tau_{z}^{(1)} \tau_{z}^{(2)}) \right], 
    \end{split}
\end{equation}
where $\delta E_z=E_z^{(2)}-E_z^{(1)}$ is the difference in Zeeman energies between the dots, $\{\sigma_j\}_{j=x,y,z}$ are the Pauli matrices in the ST basis, and $J=\frac{2 t_c^2}{U_1+\mu_1-\mu_2}+\frac{2 t_c^2}{U_2+\mu_2-\mu_1}$ is the exchange interaction using the assumption $U_i\gg\Delta_i$~\cite{kanaar_single-tone_2021}. The exchange interaction is thus controllable by the top gates via the detuning of the two dots.\par
Only the subspace corresponding to a single excited valley is relevant for this work. As a result, the zero- and doubly-excited valley states can be treated as energetically distant from the singly-excited valley subspace. Eliminating these states using another Schrieffer-Wolff transformation results in the Hamiltonian,
\begin{equation}
    \begin{split}
        H_{\text{ST},\overline{v}\text{ex}}=& \frac{J}{2}\sin \left(\frac{\delta \phi}{2}\right)^2 \sigma_{z}+\frac{\delta E_z}{2} \sigma_{x}+\frac{J}{2}\cos \left(\frac{\delta \phi}{2}\right)^2 \sigma_z\tau_x\\
        &+\Delta\left(1-\frac{J^2 \sin \left(\delta \phi \right)^2}{32\Delta_1 \Delta_2}\right) \tau_z\\
        \approx& \frac{J_s}{2} \sigma_{z}+\frac{\delta E_z}{2} \sigma_{x}+\frac{J_c}{2} \sigma_z\tau_x+\Delta \tau_z
    \end{split}
    \label{eq:HSW2}
\end{equation}
where $\{\tau_j\}_{j=x,y,z}$ are the Pauli operators acting on the subspace consisting of the valley excitation being on the left/right dot, we define the valley splitting difference $\Delta \equiv \Delta_2-\Delta_1$, and $J_s \equiv J \sin^2(\delta\phi/2), J_c \equiv J \cos^2(\delta\phi/2)$. 

In this work, we will consider $J$ on the order of 10--100~MHz~\cite{weinstein_universal_2023,wu_hamiltonian_2024,sigillito_coherent_2019,kim_approaching_2022,walelign_dynamically_2024,steinacker_violating_2024,petit_universal_2020} while $\Delta_i \gg J$, on the order of 10--100~GHz~\cite{volmer_mapping_2024,marcks_valley_2025}. Consequently, in Eq.~\eqref{eq:HSW2}, as well as in the rest of the manuscript, we ignore the term proportional to $\frac{J^2 \sin \left(\delta \phi \right)^2}{32\Delta_1 \Delta_2} \ll 1$.
In addition, since typically $\delta E_z \sim$ 1~MHz~\cite{Maune2012,walelign_dynamically_2024}, for simplicity of presentation we also neglect it in the derivations of the next section, though we include its small effect in the numerical simulations of Sec.~\ref{sec:probeQubitOperation}.\par 

\subsection{Spin-Dependent Hybridized Valley States}
\label{subsec:HV}
To simplify the Hamiltonian in Eq.~\eqref{eq:HSW2}, we move to a rotated frame where the spin-valley subspace is diagonalized. The frame transformation given by 
\begin{equation}
U_{SV}(\theta) = \exp\left(-i \frac{\theta}{2} \sigma_z \tau_y \right), \quad \theta = \tan^{-1}\left(\frac{J_c}{\Delta}\right),
\end{equation}
diagonalizes the Hamiltonian
by rotating the operators $\tau_z$ and $\sigma_z \tau_x$ into a new set of effective Pauli operators,
\seq{
T_z &= \cos \theta \, \tau_z + \sin \theta \, \sigma_z \tau_x, \\
T_x &= -\sin \theta \, \tau_z + \cos \theta \, \sigma_z \tau_x,\\
T_y &= \sigma_z \tau_y,
}
which satisfy the standard Pauli commutation relations.
In this rotated frame, since $[T_j,\sigma_z]=0\forall j=x,y,z$, one has two decoupled degrees of freedom: spin $\sigma$ and what we will refer to as a spin-dependent hybridized valley (HV) state $T$. The spin-valley coupled Hamiltonian in Eq.~\eqref{eq:HSW2} can be rewritten using the HV operators as
\begin{equation}
\label{eq:HV}
H_\mrm{SV} = \frac{J_s}{2} \sigma_z + \frac{\beta}{2} T_z,
\end{equation}
where we defined $\beta \equiv \sqrt{\Delta^2 + J_c^2}$.

\subsection{Coupling to Resonator and Master Equation}
\label{subsec:resonatorH}
We now extend the description to include coupling a detuning gate to a driven resonator. The goal is to exploit the HV degree of freedom's sensitivity to charge noise without irreversibly disturbing the spin dynamics.
We begin by describing the bare resonator with Hamiltonian
\begin{equation}
H_{\text{res}} = \omega_r a^\dagger a,
\end{equation}
where $ \omega_r $ is the resonator frequency and $ a (a^\dagger)$ is the photon annihilation (creation) operator. The voltage induced by the resonator at the detuning gate is described by
\begin{equation}
V_r = V_{rms} (a + a^\dagger),
\end{equation}
with $ V_{rms} $ the root-mean-square resonator voltage~\cite{blais_cavity_2004}. Since the exchange interaction, $ J $, is sensitive to the detuning gate voltage, we must now replace it in Eq.~\eqref{eq:HV} with the operator
\begin{equation}
\label{eq:J_resonator_coupling}
J \longrightarrow J_0 + \frac{\partial J}{\partial V_r} V_{rms} (a + a^\dagger) = J_0 + g_0 (a + a^\dagger),
\end{equation}
where we define the coupling strength $ g_0 = \frac{\partial J}{\partial V_r} V_{rms} $, and $J_0$ is a reference exchange value in the absence of resonator photons~\cite{harvey_coupling_2018}. 

Then, defining $\bar{J}_{c}=J_0\cos^2(\delta\phi/2)$, $ \theta_0 = \tan^{-1}(\bar{J}_{c}/\Delta)$, and $\omega_s=J_0\sin^2(\delta\phi/2)$, we apply $U_{SV}(\theta_0)$ to obtain the full spin-valley-resonator Hamiltonian:
\seq{
H_\mrm{SV+res} &= U_{SV}(\theta_0) \left(H_{\text{SV}} + H_\mrm{res} \right) U_{SV}(\theta_0)^\dagger \\
&= \frac{\omega_s}{2}\sigma_z +  \omega_r a^\dagger a + \frac{\beta_0}{2} T_z \\
& + \frac{g_0}{2} \cos \theta_0 T_x (a + a^\dagger) + \frac{g_0}{2} \sin \theta_0 T_z (a + a^\dagger),
}
where $\beta_0=\sqrt{\Delta^2+\bar{J}_{c}^2}$.
The last term represents longitudinal coupling and leads to a small energy shift. For small $ g_0 $ and large valley splitting difference $\Delta \gg J_0$, it can be neglected, leaving the transverse coupling term
\begin{equation}
\frac{g}{2} T_x (a + a^\dagger), \quad g = g_0 \cos \theta_0,
\end{equation}
which, as we will show below, leads to a Jaynes-Cummings-type of interaction~\cite{jaynes1963}.

We next include resonator driving by an external probe field at frequency $ \omega_p $~\cite{mutter_fingerprints_2022}, modeled by
\seq{
H_d = \epsilon_d (a e^{-i\omega_p t} + a^\dagger e^{i\omega_p t}),
}
and move to a rotating frame at frequency $ \omega_p $ using the transformation $U_\mrm{RWA} = \exp[i \omega_p t (T_z + a^\dagger a)]$, which after applying the rotating wave approximation (RWA) yields the RWA Hamiltonian:
\begin{multline}
H_{\text{RWA}} = U_\mrm{RWA}\left(H_\mrm{SV+res} + H_d \right) U_\mrm{RWA}^\dag - i \hbar U_\mrm{RWA} \dot{U}_\mrm{RWA}^\dag
\nonumber \\
=\frac{\omega_s}{2}\sz + (\omega_r - \omega_p)a^\dagger a + (\beta_0 - \omega_p)T_z \nonumber \\
+ g (a^\dagger T_- + a T_+) + \epsilon_d(a + a^\dagger),
\end{multline}
where $ T_\pm = (T_x \pm i T_y)/2 $. In the dispersive regime where the resonator-HV detuning $\Delta_{rv} \equiv \beta_0 - \omega_r \gg g$, we can adiabatically eliminate the cavity to obtain the effective dispersive Hamiltonian~\cite{Blais2021}:
\begin{equation}
\label{eq:H_disp}
H_\text{disp} = \frac{\omega_s}{2} \sigma_z + \delta_p a^\dagger a + \frac{\delta_\beta + \chi a^\dagger a}{2} T_z+ \epsilon_d(a + a^\dagger),
\end{equation}
with dispersive shift $ \chi = g^2/\Delta_{rv} $, and $\delta_p=\omega_r-\omega_p$ and $\delta_\beta=\beta_0-\omega_p$.
This interaction leads to a photon-number-dependent shift of the HV frequency while leaving the spin frequency unaffected. 
Note that the validity of the dispersive Hamiltonian in Eq.~\eqref{eq:H_disp} depends on the system being weakly driven such that the number of photons is much smaller than the critical number of photons
\begin{equation}
    n_{\text{crit}}=(\Delta_{rv}/2g)^2.
\end{equation}

Finally, including resonator decay at rate $ \kappa $, the system dynamics are governed by the master equation
\begin{equation}
\label{eq:LME}
\frac{\partial \rho}{\partial t} = -i[H_{\text{disp}}, \rho] + \frac{\kappa}{2} (2a \rho a^\dagger - a^\dagger a \rho - \rho a^\dagger a).
\end{equation}
Although valley relaxation would take the system outside the computational subspace, the valley relaxation rate $\sim 10–100$~kHz~\cite{penthorn_direct_2020} is slow compared to the resonator dynamics, so we can neglect it throughout this work.

\subsection{Resonator Dynamics and Homodyne Measurement}
\label{subsec:res_dyn_hom_meas}
When the resonator is in a coherent  state, Eq.~\eqref{eq:LME} can be solved assuming the HV is in a ground ($g$) or excited ($e$) state, resulting in field amplitudes given by $\alpha_{g,e}(t)=\langle a(t)\rangle_{g,e}$ whose time evolutions are dictated by~\cite{frisk_kockum_undoing_2012,Gambetta2006}
\seq{
\label{eq:alpha_ss}
\dot{\alpha}_{e,g} &= -i \epsilon_d - i (\delta_p\pm\chi)\alpha_{e,g} - \frac{\kappa}{2} \alpha_{e,g}.
}
Assuming that the resonator is initially empty and the probe field is turned on at $t=0$, the solution to these equations is given by
\seq{
\alpha_{e,g}(t) &=  \frac{\epsilon_d}{(\delta_p\pm\chi) + i \kappa/2}\l(1 - e^{-\left(\frac{\kappa}{2}-i(\delta_p\pm\chi)\right)t}\r),
}
which approaches its steady state with equilibration time $2\kappa^{-1}$.
In particular, note that these equations possess steady state solutions given by
\seq{
\alpha_{e,g}^{ss} = \frac{-\epsilon_d}{\delta_p\pm\chi-i\kappa/2},
}
to which the pointer states converge exponentially with a rate $\kappa/2$. 
From Eq.~(\ref{eq:alpha_ss}) we can see that
\seq{
\label{eq:imrealpha}
\mrm{Im}[\alpha_{e,g}^{ss}] &= \frac{- \epsilon_d \kappa/2}{(\delta_p\pm\chi)^2+\kappa^2/4},\\
\mrm{Re}[\alpha_{e,g}^{ss}] &=\frac{-\epsilon_d (\delta_p\pm\chi)}{(\delta_p\pm\chi)^2+\kappa^2/4}.
}

For a homodyne measurement with local oscillator (LO) phase $\phi$, the operator $X_\phi=a e^{-i\phi}+a^\dagger e^{i\phi}$ is measured. Then the output signal becomes proportional to the homodyne current~\cite{frisk_kockum_undoing_2012}
\begin{equation}
\label{eq:jt_def}
j(t) = \sqrt{\kappa} \braket{X_{\phi}(t)} + \xi(t),
\end{equation}
where we have assumed unit measurement efficiency, and $\xi(t)$ represents a white shot noise contribution to the measurement, satisfying $\mathbb{E}[\xi(t)]=0$ and $\mathbb{E}[\xi(t)\xi(t')]=\delta(t-t')$. Here, the $\delta(t)$ function is normalized such that $\mathbb{E}[\xi(t)^2]=dt^{-1}$, where $dt$ is the time interval setting the measurement rate.

\section{Charge-Noise Detection Protocol}
\label{sec:protocol}

In this section, we describe the measurement protocol introduced in this work. 
To activate the probe, the resonator is driven with $\epsilon_d>0$, populating it with roughly 
\begin{equation}\label{eq:n}
    \langle a^\dagger a \rangle \approx |\alpha_g^{ss}|^2 = \frac{\epsilon_d^2}{(\delta_p-\chi)^2+\kappa^2/4}
\end{equation}
steady-state photons. Throughout this section, we assume that the HV is initialized to its ground state and that, due to the large $\Delta$, it remains close to its ground state throughout the evolution.

\subsection{Signal}
\label{subsec:signal}
Charge noise modifies the valley splitting difference and exchange coupling to $\Delta'= \Delta + \delta\Delta$ and $J'=J+\delta J$, respectively, which in turns modifies the coupling strength $g'=g+\delta g$, detuning $\Delta_{rv}'=\Delta_{rv}+\delta\Delta_{rv}$, and dispersive shift $\chi \rightarrow \chi'=g'^2/\Delta_{rv}' = \chi + \delta \chi$. Moreover, since $\Delta\gg J$, we have that $\delta\Delta_{rv}=\Delta \delta\Delta/\sqrt{\bar{J}_c^2+\Delta^2} = \delta\Delta ( 1 + O(J^2/\Delta^2)) \approx \delta\Delta$. Assuming that the noise effect is weak, namely $|\delta\Delta/\Delta|,|\delta g/g| \ll 1$, the leading order expression for the change in dispersive shift is 
\begin{equation}\label{eq:dchi}
    \delta\chi=\chi\l(2 \frac{\delta g}{g} -\frac{\delta\Delta}{\Delta_{rv}}\r),
\end{equation}
producing a detectable change in the homodyne voltage $V(t) = G j(t)$ where $G$ is the total amplifier chain gain. 
In addition, the charge noise shift also directly affects $\beta_0$ through both $\Delta$ and $J$, leading to an unobservable change in the HV energy. 

In order to optimize the effect of the noise on the voltage, we choose $\delta_p=0$ and take $d\braket{X_{\phi=\phi_\mrm{opt}}}/d\chi=0$. From here we find the optimal LO phase 
\begin{equation}
\label{eq:phi_opt}
\phi_\mrm{opt}(\chi_0)=\arctan\l(\frac{\kappa\chi_0}{\chi_0^2-\kappa^2/4}\r),
\end{equation} 
where $\chi_0$ is the baseline, unperturbed value of $\chi$. This choice produces the optimal signal. 
Note that, in the limit $\chi/\kappa\rightarrow0$, Eq.~\eqref{eq:phi_opt} reduces to $\phi_\mrm{opt}\rightarrow0$, or equivalently, measuring $\mrm{Re}\braket{a}$.
We then obtain the deterministic part of the stationary homodyne current at that phase for arbitrary $\chi$ to be
\begin{multline}
\label{eq:jdet}
    j_\mrm{det}(\chi,\chi_0) = 2\sqrt{\kappa}\left( \cos\phi_\mrm{opt}(\chi_0) \re\alpha_g^{ss}(\chi) \right.
    \\
    + \left. \sin\phi_\mrm{opt}(\chi_0) \im\alpha_g^{ss}(\chi)\right),
\end{multline}
where the dependence of $\alpha_g^{ss}$ on $\chi$ was made explicit to reinforce that this is the only part of the current that is perturbed by noise; $\phi_\mrm{opt}$ is calibrated to the baseline value of $\chi_0$ when preparing the LO and is fixed throughout.
In addition, the unperturbed homodyne current is $j_\mrm{det}(\chi_0,\chi_0)=2\sqrt{\kappa}\chi_0\epsilon_d/(\kappa^2/4+\chi_0^2)$.
Any change $\delta\chi$ in the dispersive shift due to charge noise will induce a current change at LO phase $\phi_\mrm{opt}(\chi_0)$ of
\begin{align}
\label{eq:dj_det}
\delta j_\mrm{det}(\chi_0,\delta\chi) &= |j_\mrm{det}(\chi_0+\delta\chi,\chi_0) - j_\mrm{det}(\chi_0,\chi_0)| \\
\label{eq:dj_det_approx}
&\approx \l| \frac{d j_\mrm{det} (\chi,\chi_0)}{d \chi} \r|_{\chi=\chi_0} \delta \chi = \frac{2\epsilon_d \sqrt{\kappa}}{\kappa^2/4+\chi_0^2} \delta \chi. 
\end{align}

We define the stochastic signal by comparing the measured current at the optimal phase against the baseline value,
\seq{
s(t) = j(t) - j_\mrm{det}(\chi_0,\chi_0).
}
Denoting by $t_n$ the time when the charge noise event occurs, note that $\mathbb{E}[s(t< t_n)]=0$ and $\mathbb{E}[s(t\gg t_n)]=\delta j_\mrm{det}(\chi_0,\delta\chi,\chi_0)$, where $\mathbb{E}[\cdot]$ denotes the average over $\xi(t)$.
Furthermore, in practice the measurements will be carried out over times significantly longer than $1/\kappa$, so we ignore the transients and work directly with the steady-state values.

\subsection{Signal-to-Noise Ratio}
\label{subsec:SNR}

The signal after a charge noise event has occurred is given by $s(t\gg \kappa^{-1})\approx \delta j_\mrm{det}(\chi_0,\delta\chi) + \xi(t)$, from which it is clear that the capability to detect an event depends on the relative magnitudes of the shift $\delta j_\mrm{det}(\chi_0,\delta\chi)$ and the shot noise $\xi(t)$.
Given a measurement time $t_m$, the signal-to-noise ratio (SNR) is defined using the integrated signal, by 
\seq{
\SNR(t_m) &= \frac{\l| \int_0^{t_m}  \delta j_\mrm{det}(\chi_0,\delta\chi) dt \r|}{\sqrt{\int_0^{t_m}\int_0^{t_m} \mathbb{E}[\xi(t)\xi(t')] dt dt'} }\\
&\approx \frac{ 2 \epsilon_d \sqrt{\kappa} }{\kappa^{2}/4+\chi_0^2} \sqrt{t_m} |\delta\chi|,
}
where we used $\int_0^{t_m} \mathbb{E}[\xi(t)\xi(t')] dt = 1 \forall t'\in[0,t_m]$.

More explicitly, when the readout chain setup includes a quantum-limited amplifier, such as a Josephson parametric amplifier (JPA), the quantum shot noise will be the dominant source of uncertainty in determining $j_\mrm{det}$ from $s(t)$. We solve $\SNR(t_m^\mrm{shot})=1$ for $t_m^\mrm{shot}$ to obtain the measurement time needed to detect a dispersive shift of minimal resolution $\delta\chi_\mrm{min}$:
\seq{
\label{eq:tm_vs_n}
t_m^\mrm{shot} &= \frac{\kappa^2/4+\chi_0^2}{4 \,\kappa\, \delta\chi_\mrm{min}^2 \braket{n}},
}
where we used Eq.~\eqref{eq:n} for the number of photons, $\braket{n}=\braket{a^\dagger a}$

Although on the one hand it is desirable to use a large number of photons to decrease the measurement time, to ensure the validity of the dispersive approximation it is necessary to remain within the dispersive regime, which requires $\braket{n} \ll n_{\text{crit}}$.

We now find typical values of $\delta \chi$ by relating it to the experimentally measured quantities. Charge noise measurements are often expressed in terms of the average chemical potential shift of a dot, $\delta\mu$. When that shift is due to a power spectral density 
\begin{equation}
    S(f) = \frac{A_0^2}{f}
\end{equation}
measured over a time, $t_m$, one has
\begin{equation}
\label{eq:dmu}
    \delta \mu = A_0 \sqrt{\log(f_\text{cutoff} t_m}),
\end{equation}
where $f_\text{cutoff}$ is the noise cutoff frequency. 
Experimentally obtained charge noise parameters $A_0$ and $f_\mrm{cutoff}$ are shown in Table~\ref{tab:params}. For our protocol, we consider measurement times in the range of $1\,\mu$s to 1~ms: shorter than estimated valley relaxation times of $\sim 10\,$ms~\cite{penthorn_direct_2020} and much longer than a typical measurement rate $dt\approx 10\,$ns to ensure sufficient shot noise averaging~\cite{Walter2017}. In this setting, charge noise would induce chemical potential shifts in the range of $\delta\mu \approx 2$--$6\,\mu$eV, respectively.

The corresponding fluctuations in $J$ can be estimated through its measured dependence on a gate voltage as well as the known proportionality constant (or ``lever arm") between that gate voltage and the chemical potential, $\delta \mu = \alpha \delta \epsilon$, where $\delta \epsilon$ is the corresponding ``gate-referred" voltage fluctuation \cite{reed_reduced_2016}. Thus, for $\delta\mu\approx4\,\mu$eV, with a lever arm of 0.2 eV/V~\cite{cai_coherent_2023}, the corresponding shift in detuning gate voltage is $\delta\epsilon\approx 20\,\mu$V. 
For the device of Ref.~\cite{cai_coherent_2023}, in which detailed data of the detuning dependence of exchange is given, we find that the relative shift in exchange
when operating in the large-$J$ regime at $\epsilon=1.1\,$mV is as large as $\delta J/J \approx 3\%$, and the sensitivity becomes $\partial J/\partial \epsilon = \partial J/\partial V \approx0.5\,$MHz$/\mu$eV.

Meanwhile,  
\begin{equation}
    \delta \Delta = (\vec{\nabla}_F \Delta_{2}-\vec{\nabla}_F \Delta_{1}) \cdot \delta \vec{F} \approx \vec{\nabla}_F \Delta_{1/2} \cdot \delta \vec{F},
\end{equation}
where $\delta \vec{F}$ is the expected charge-noise-induced fluctuation in the electric field and $\vec{\nabla}_F \Delta_{1/2}$ is the sensitivity of the individual dots' valley splitting to the electric field. Each electron samples different alloy disorder~\cite{losert_practical_2023,marcks_valley_2025}, so the fluctuations in $\Delta_1$ and $\Delta_2$ are uncorrelated and a conservative estimate of the fluctuation in the difference is the characteristic size of the individual fluctuations.
The average value of $|\vec{\nabla}_F \Delta_{1/2}|$ for a pure silicon quantum well with a realistic smearing of the interface due to alloy disorder to a width of $w=0.5$~nm~\cite{esposti_low_2023} has been calculated to be approximately 0.2~e~$\cdot$~nm \cite{kanaar_proposed_2024}, primarily in the lateral direction. The electric field fluctuation is estimated as
\begin{equation}
    |\delta F|=\frac{\delta \epsilon}{l}=\frac{\delta \mu}{\alpha l}.
\end{equation}
We have assumed that the voltage induced by charge noise varies approximately inversely with the distance between the dot and the gate, $l$. 
So 
\seq{
\label{eq:dDelta_vs_deps}
\delta\Delta \approx \delta \epsilon \, d / l.
}

Using Eq.~\eqref{eq:dDelta_vs_deps} with $|\vec{\nabla}_F \Delta_{1/2}|=0.2\,e \cdot $nm, we find that the change in valley splitting difference is $\delta \Delta \approx 0.1\,\mu$eV. With typical valley splitting differences being in the order of 10--100$\,\mu$eV, we estimate that the fluctuations in the range  $\delta \Delta/\Delta\sim 0.1-1\%$. 
In what follows, we estimate the magnitude of a typical fluctuation in $\chi$ by 
\seq{
\label{eq:dchi_est}
\delta\chi_\mrm{est} = \chi_0 \sqrt{\l(2\frac{\delta J}{J}\r)^2+\l(\frac{\delta \Delta}{\Delta_{rv}}\r)^2}.
}
Note that, in addition, the change $\delta\chi$ depends on the measurement time through $\delta\mu$ (see Eq.~\eqref{eq:dmu}), which makes the inequality \eqref{eq:tm_vs_n} nonlinear in $t_m$. However, since the logarithmic dependence is so weak, in practice we can safely ignore it in the following and simply take a uniform shift distribution in chemical potential over the measurement times of interest.
If a quantum-limited amplifier is used the dominant source of measurement noise is shot noise and, using the relations  $g \approx \frac{dJ}{dV_r} V_{rms}$, $V_{rms}=\omega_r \sqrt{Z \hbar}$, and $\kappa = \omega_r / 2Q$, where $Z$ is the resonator impedance and $Q$ is the quality factor, the measurement time can be written in terms of resonator parameters and the effect of charge noise on the dispersive shift $\delta \chi/\chi_0$:
\begin{equation}
    t_m^{\mrm{shot}} \approx \left(8 \frac{\braket{n}}{n_\mrm{crit}} \left(\frac{\delta \chi}{\chi_0} \frac{dJ}{dV}\right)^2\hbar\omega_r Q Z    \right)^{-1}.
\end{equation}
We choose representative values for the parameters listed in Table~\ref{tab:params}, which yields a measurement time of 13~$\mu$s.The photon number for the probe, $\braket{n}/n_\mrm{crit}$, is set to 0.025 due to simulation constraints, but in practice could be increased to 0.1 without violating the dispersive regime approximations.   
This formulation also allows targeting a specific noise sensitivity or measurement time by appropriately choosing the resonator parameters $Q$, $\omega_r$, and $Z$.\par
If a quantum-limited amplifier is not used, the dominant source of noise comes from the cryogenic amplifier. The full derivation of the SNR for this case is shown in Appendix \ref{app:SNR}. Replacing the signal voltage using $V_0=\sqrt{R \hbar \omega_r }\, \delta j_\mrm{det}/2$ and  applying the same substitutions defined above, the measurement time for unity SNR in the presence of thermal noise becomes:
\begin{equation}
\label{eq:tm_therm}
    t_m^{\mrm{cryo}} \approx \frac{7 k_B T}{\hbar\omega_r} \times t_m^{\mrm{shot}}.
\end{equation}
Using the parameters in Table~\ref{tab:params}, we find a measurement time of 9~ms. However, decreasing the stray capacitance by a factor of 10, for example, would increase each of $\omega_r$, $Q$, and $Z$ by a factor of $\sim 3$ and decrease the measurement time to $\sim 90~\mu$s. Effective operation of the probe without a quantum-limited amplifier thus depends sensitively on the resonator parameters and the exchange sensitivity to voltage, $\frac{dJ}{dV}$, all of which can be engineered through device design.
\begin{table}[t!]
    \centering
    \begin{tabular}{l l l}\hline
       Parameter  & Symbol& Value \\ \hline
       Charge noise strength &$A_0$ & 1.0 $\mu$eV~\cite{connors_low-frequency_2019,struck_low-frequency_2020,freeman_comparison_2016,connors_charge-noise_2022,paquelet_wuetz_reducing_2023, esposti_low_2023}\\
       High frequency noise cutoff &$f_{\text{cutoff}}$ &100 MHz~\cite{connors_charge-noise_2022}\\
       Lever arm & $\alpha$ & $0.2\,$eV/V~\cite{cai_coherent_2023}\\
       Dot distance from gate & $l$& 100 nm \cite{yoneda_quantum-dot_2018}\\
       Resonator frequency  & $\omega_r$ & 5 GHz~\cite{Kandel2024,Koolstra2025,Borjans2021} \\
       Resonator quality factor&  $Q$& $2500$~\cite{Kandel2024,Koolstra2025,Borjans2021,samkharadze_high-kinetic-inductance_2016}\\
       Resonator impedance &$Z$  &$10^4$ $\Omega$~\cite{Kandel2024,Koolstra2025,Borjans2021,samkharadze_high-kinetic-inductance_2016} \\
       Transmission line resistance & $R$ & 50 $\Omega$\\
       Valley splitting difference & $\Delta$ & 20 $\mu$eV~\cite{volmer_mapping_2024} \\
       Cryo amplifier temperature & $T$ & 4\,K \\
       Number of photons & $\braket{n}/n_\mrm{crit}$ & 0.025
    \end{tabular}
    \caption{Parameter values used to calculate the measurement times for unity SNR.}
    \label{tab:params}
\end{table}

\section{Impact on Qubit Evolution and State Restoration}
\label{sec:probeQubitOperation}
The previous section shows that a measurement of a fluctuation within a reasonable time is possible for a high-quality resonator. For the probe to be used to assist in qubit operation there are two further considerations: how the probe operation influences the dynamics of the ST qubit, and how the information obtained from the probe may be used to improve qubit operation after a fluctuation is detected. In this section, both those issues will be discussed.

\subsection{Response to fluctuation}
\label{subsec:response}
When a shift in exchange occurs, the probe signal will change accordingly. The map between the exchange and the detuning voltage is usually characterized before qubit operation and is an injective function, meaning that knowing the shift in detuning voltage allows one to know the new exchange and fix it if desired. For small shifts, the signal change is approximately proportional to the shift in detuning voltage, allowing for real-time compensation by adjusting the detuning voltage back to the desired point in exchange.\par

\begin{figure*}[t!]
    \centering
    \includegraphics[width=\linewidth]{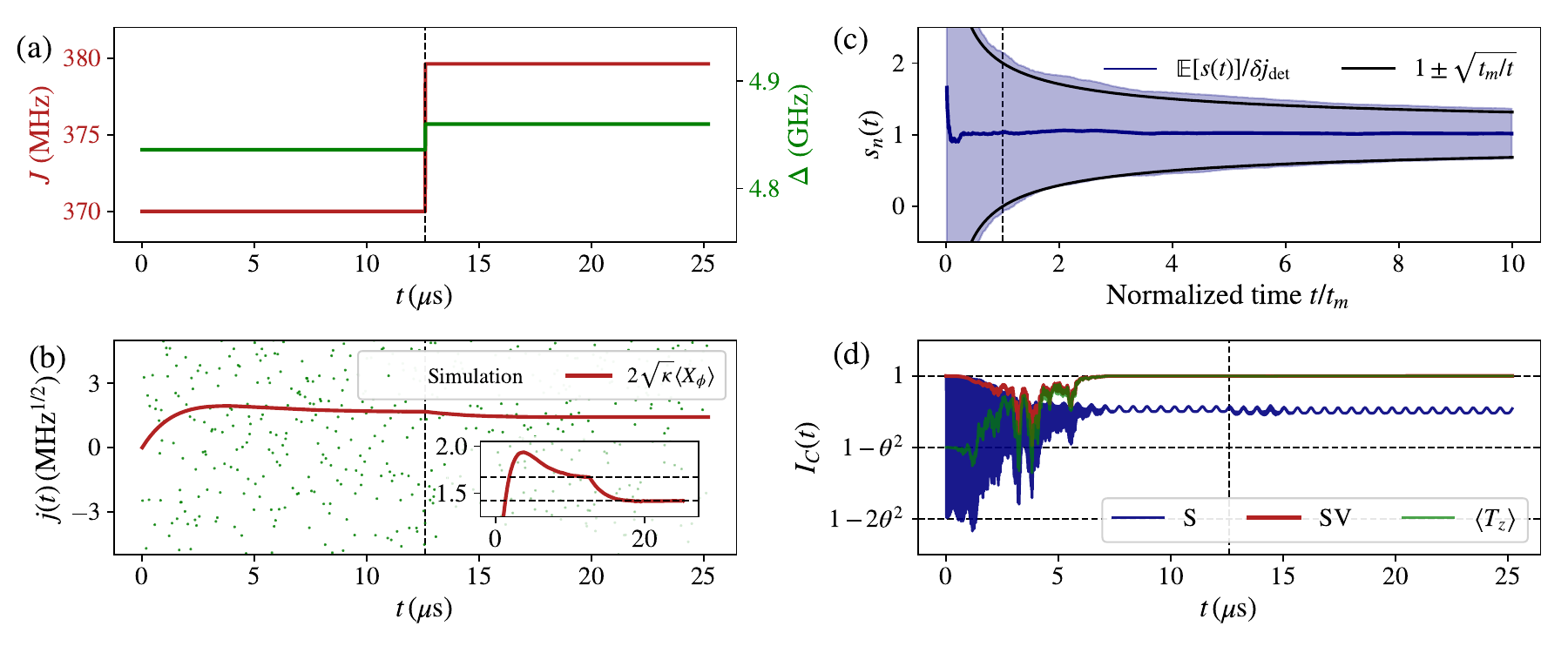}
    
    \captionsetup{justification=centerlast,singlelinecheck=false}
    \caption{Example of charge noise detection scheme, simulating Eq.~\eqref{eq:LME} via stochastic master equation using Qutip~\cite{qutip}. The parameters used in simulation are $\omega_s=10\,\mrm{MHz}, \bar{J}_{c}=370\,\mrm{MHz},\epsilon_d=1\,\mrm{MHz},\kappa=10\,\mrm{MHz}, \delta E_z=2\,$MHz and $dt=1\,$ns, in addition to those shown in Table~\ref{tab:params}. The charge-noise effect was set to $\delta \chi/\chi=0.13$, following the discussion in Sec.~\ref{subsec:SNR} and Eq.~\eqref{eq:dchi_est}. 
    The vertical dashed line marks the charge-noise event time $t_n=13\,\mu$s. 
    (a) Exchange and valley splitting difference as function of time, where both quantities shift at $t_n$.
    (b) Homodyne current (green) with the underlying deterministic contribution of the signal (red). The inset shows a zoomed-in view. 
    (c) Normalized signal Eq.~\eqref{eq:norm_signal}, averaged over 1000 different realizations. (d) Purity over each subsystem partition: spin only (S) and combined spin-valley (SV). Expectation value of the HV $T_z$ operator was added for reference. }
    \label{fig:results}
\end{figure*}

The shift in signal is shown in Fig.~\ref{fig:results}, where the system is initialized with an empty cavity, and the HV is initially in the ground state $\ket{0,-}$; at $t=0$, a pulse is applied to take the spin to the $(\ket{0}+\ket{1})/\sqrt{2}$ state.
In Fig.~\ref{fig:results}(a), we show the exchange and valley splitting difference as function of time. At time $t_n$, a charge-noise event occurs that shifts both by $\delta \Delta = 24\,$MHz and $\delta J = 10\,$MHz, respectively. 
In Fig.~\ref{fig:results}(b), we present the stochastic homodyne current obtained from monitoring the resonator in green, as well as the underlying deterministic component $j_\mrm{det}$. 

At time $t_n$ the simulated charge noise event shifts the value of $\chi$ by 13\%. The evolution of the system is shown for another $13\mu$s after the noise event.

We confirm numerically the measurement time estimate in Eq.~\eqref{eq:tm_vs_n}. In Fig.~\ref{fig:results}(c), we show the normalized signal
\seq{
\label{eq:norm_signal}
s_n(t) = \frac{\mathbb{E}[s(t)]}{\delta j_\mrm{det}},
}
averaged over 100 realizations at each time $t$. The shaded areas denote the standard deviations over these realizations. Normalizing the horizontal axis by the measurement time $t_m$ for $\mrm{SNR}=1$, we can see that the average normalized signal crosses the $100$\% line near $t=t_m$, namely the points $s_n(t_m)\approx 1\pm 1$.
For longer measurement times, the uncertainty decreases as $\sqrt{t_m/t}$, and the presence of a shift can be determined with greater certainty. 
This property shows that the occurrence of a noise event can be determined within $O(t_m)$ albeit with a large uncertainty in the value of $\delta\chi$, and that measuring for longer times $t\gg t_m$ provides increasingly accurate estimates of the magnitude of the shift.\par
\begin{figure}[ht!]
    \centering
    \includegraphics[width=\linewidth]{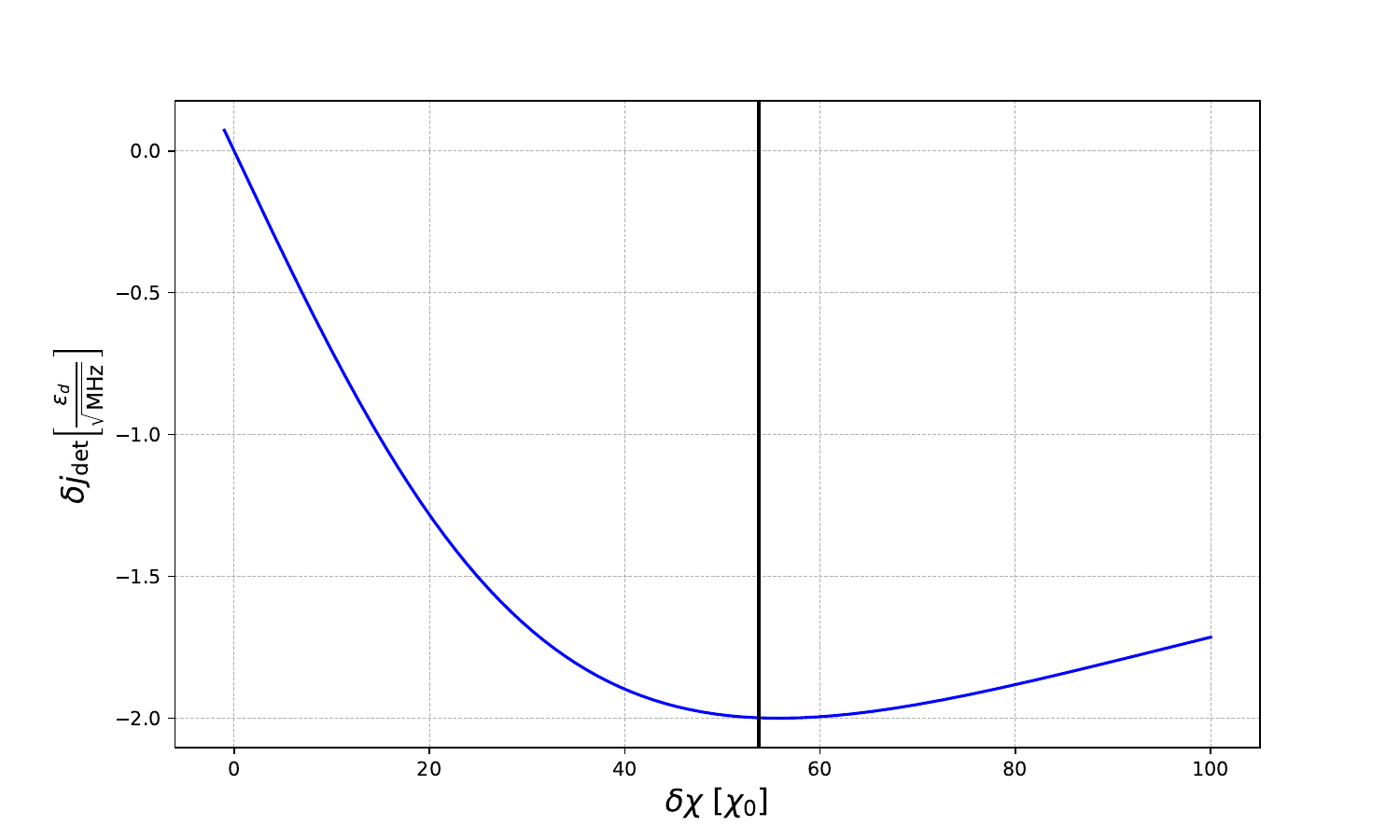}
    \captionsetup{justification=centerlast,singlelinecheck=false}
    \caption{The signal strength, $\delta j_\mrm{det}$,  as a function of change in dispersive shift $\delta \chi$ using the parameters of Table \ref{tab:params}. The thick vertical line indicates $\delta\chi_\mrm{max}^+$.}
    \label{fig:deltajdet}
\end{figure}
A potential complication arises from the fact that the probe signal, $\delta j_\mrm{det}$, is nonmonotonic in the change of dispersive shift, $\delta\chi$, as shown in Eq.~\eqref{eq:dj_det}. As a result, a given value of $\delta j_\mrm{det}$ corresponds to two values of $\delta\chi$ making the inversion of the signal ambiguous. This behavior is illustrated in Fig.\ref{fig:deltajdet}, where $\delta j_\mrm{det}$ is plotted as a function of $\delta \chi$ with the values in Table.~\ref{tab:params}. The function remains single-valued only in the restricted domain of $\delta\chi \in [\delta\chi_\mrm{max}^-,\delta\chi_\mrm{max}^+]$, $\delta\chi_\mrm{max}^\pm=(\kappa^2+4\chi_0^2)/(\pm 2\kappa-4\chi_0)$, though in practice $\delta \chi $ cannot be smaller than $-\chi_0$ since $\chi$ is always positive.
Because most charge noise-induced fluctuations are expected to fall within the monotonic domain, it is typically reasonable to assume the smaller of the two possible $\delta \chi$ values and apply a corresponding correction to the top gate voltages to restore the exchange interaction to its baseline value. If the homodyne current does not return to its baseline after this adjustment and a sufficient equilibration time, one can then infer that a larger shift occurred and apply an updated correction. \par
If the valley excitation decays during probing, the Hamiltonian no longer contains tranverse coupling to the resonator. Thus, a valley decay would cause $j_{\mrm{det}}$ to go to zero, equivalent to the effect of an exchange fluctuation such that $\delta\chi =\chi_0(\kappa^2+4\chi_0^2)/(\kappa-4\chi_0)$. One could distinguish it from such an exchange fluctuation by checking whether the signal can be restored with the previously described procedure. If the signal does not respond, it is confirmed to be a decay event. At that point the valley excitation can be repumped by sweeping the detuning through the spin-valley avoided crossings as described in Sec.~\ref{sec:Model}.\par 
An alternative approach to this active feedback approach is to passively track the fluctuations in the exchange interaction and use feedforward to adjust the subsequent operations to compensate. For instance, in the case of a single-qubit, fluctuations at large exchange result in extra $Z$ rotations, which, in a resonant operation mode \cite{takeda_resonantly_2020}, can be compensated by virtual $Z$ gates \cite{mckay_efficient_2017}. The advantage of this approach is that by eliminating the active recalibration steps it may allow for longer integration times before applying corrections, increasing confidence in the accuracy of the inferred shift in $J$. However, a potential drawback is that this imposes constraints on the circuit so as to ensure that all entangling operations are performed via controlled phase gates that commute with local $Z$ rotations, which may be cumbersome.

\subsection{Spin evolution negligibly affected by probe}
\label{subsec:HV_evo}
In this section, we show that the presence of the valley-probe setting does not limit the parallel operability the qubit. Going back to the Hamiltonian introduced in Eq.~\eqref{eq:HV}, we note that since $[T_z,\sigma_z]=0$, the Hamiltonian can be simplified by going to a frame rotating with $J_{s} \sigma_z/2$ and we can focus strictly on the HV dynamics in this section. 
We assume the HV is initially in its ground state, 
\seq{
\ket{\psi_g} = \ket{0}\l[\cos\l(\frac{\theta}{2}\r)\ket{v_-}+\sin\l(\frac{\theta}{2}\r)\ket{v_+}\r],
}
where $\ket{0/1}$ denote the ground ($0$) and excited $(1)$ states of the spin, $\ket{v_\pm}$ denote the ground ($-$) and excited $(+)$ states of the valleys, and it is easy to show that $T_z\ket{\psi_g}=-\ket{\psi_g}$.
Note that, since $\theta\ll1$, this state is close to the decoupled valley-spin ground state, namely $\ket{\psi_g}\approx \ket{0}\ket{v_-}$.
It is useful to write the other $T_z$ eigenstates
\seq{
\ket{0,+} &= \ket{0}\l[\sin\l(\frac{\theta}{2}\r)\ket{v_-}-\cos\l(\frac{\theta}{2}\r)\ket{v_+}\r], \\
\ket{1,-} &= \ket{1}\l[\cos\l(\frac{\theta}{2}\r)\ket{v_-}-\sin\l(\frac{\theta}{2}\r)\ket{v_+}\r], \\
\ket{1,+} &= \ket{1}\l[\sin\l(\frac{\theta}{2}\r)\ket{v_-}+\cos\l(\frac{\theta}{2}\r)\ket{v_+}\r],
}
where the state $\ket{s,u}$ denotes that the ST spin degree of freedom is in the $\ket{s}$ state, and the eigenstate has $T_z$ eigenvalue sign $u=\pm1$, i.e., $T_z\ket{s,u}=u\ket{s,u}$.
In addition, $\ket{\psi_g}=\ket{0,-}$ since it is also the spin ground state, and is thus the collective ground state.
Consequently, the four eigenvalues of $T_z$ comprise two degenerate pairs, with each pair of degenerate eigenstates consisting of one ground and one excited spin state.

To study the impact of the Hamiltonian Eq.~\eqref{eq:HV} on the spin evolution, we prepare the spin in an arbitrary state $a\ket{0}+b\ket{1}$, with $\sqrt{|a|^2+|b|^2}=1$. In terms of the eigenstates of $H_\mrm{SV}$, this state reads
\seq{
\label{eq:HV_psi0}
\ket{\psi(0)} &= (a\ket{0}+b\ket{1})\l[\cos\l(\frac{\theta}{2}\r)\ket{v_-}+\sin\l(\frac{\theta}{2}\r)\ket{v_+}\r] \\
&= a \ket{0,-}+b\cos\theta\ket{1,-}+b\sin\theta\ket{1,+}.
}
Under the Hamiltonian Eq.~\eqref{eq:HV}, this state evolves (again, up to an overall rotation by $\exp \l(-i J_s t \sigma_z/2\r)$ that is independent of the valley and resonator degrees of freedom and not relevant to the present consideration) as
\seq{
\label{eq:psi_sv_t}
\ket{\psi(t)} &= e^{-i H_\mrm{SV} t } \ket{\psi(0)} \\
&= e^{i\beta t} (a \ket{0,-}+b\cos\theta\ket{1,-})\\
&+ e^{-i\beta t}b\sin\theta\ket{1,+}.
}
Then it is easy to compute the survival probability of the prepared state, 
\seq{
p(t;a,b) &= |\bra{\psi(0)} \psi(t) \rangle |^2 \\
&= 1-4 |b|^2 (|a|^2+|b|^2\cos^2\theta)\sin^2(\theta)\sin^2(\beta t/2)\\
&\approx 1 - 4 |b|^2 \theta^2 \sin^2\l(\frac{\beta t}{2}\r),
}
from where we can see that $p(t;a,b)=1-O(\theta^2)$.
Thus, the coupling with the valleys is restricted to a $\theta^2\ll1$ contribution. In particular for the parameters of Table~\ref{tab:params}, we have $\theta^2=4\cdot 10^{-4}$ and consequently ignore this contribution in this work.

Lastly, we note that the measurement with local oscillator phase of $\phi=\phi_\mrm{opt}$ will in general be distinct from $\pi/2$, and thus it will  distinguish between the HV states, leading to state collapse~\cite{frisk_kockum_undoing_2012,Blais2021}. However, as long as the valley is near its ground state when the measurement starts, the collapse will be towards states with eigenvalue $-1$. 
As can be seen from Eq.~\eqref{eq:psi_sv_t}, the contribution of a state with eigenvalue $+1$ is of order $\sim \theta$, and thus we can see that the state collapse will not appreciably affect the spin degree of freedom. This is confirmed numerically in simulation of the stochastic master equation, shown in Fig.~\ref{fig:results}(c).
Here, we show the purity of different subsystems: spin only (S) and combined spin-valley (SV).
The purities are computed by evolving the full density matrix $\rho(t)$, and taking the partial trace $\rho_C(t)=\tr_{\bar{C}}[\rho(t)]$ over the complement of subsystem $C\in\{\mrm{S,SV}\}$, denoted by $\bar{C}$. Then, the purities are defined by $I_C=\tr[\rho_C^2(t)]$.

As shown in Fig.~\ref{fig:results}(c), as the resonator is populated, both the purities go through a transition until reaching equilibrium with characteristic time $2\kappa^{-1}$. In particular, it is worth noting that the SV purity converges to 1, while the S purity go to $1-\theta^2/2$. This can be understood by noting that $\braket{T_z}$ (shown in green) is forced to converge to an eigenstate due to the measurement-induced collapse, observed in $\braket{T_z}\rightarrow1$. Note that $\braket{T_z}$ starts at $1-\theta^2$ due to the state being initialized in the spin $\ket{+}$ state, consistent with Eq.~\eqref{eq:HV_psi0}.

At time $t_n$, the charge-noise event leads to a new transitory period.
After equilibration, the HV returns to unit purity again due to state collapse. However, due to the change in $\beta_0$ induced by charge-noise, the ground state has changed, and $\braket{T_z}$ does not return to 1. Note that if the operator $\braket{T_z}$ with the new $\theta$ was used instead, this expectation value would return to 1.
Lastly, even though all these effects may be readily understood in terms of the state evolution and collapse due to measurement, they are all within $O(\theta^2)$. Since in this work we consider the regime where $J\ll\Delta$, this allows us to ignore the influence of the valley and resonator on the spin.

\section{Discussion and Conclusion}\label{sec:Summary}
In this work, we have introduced a method for real-time, in-situ detection of charge noise using a singlet-triplet qubit with one of the two electrons initialized in an excited valley state and coupled to a resonator. The singly-excited-valley states are coupled via the exchange interaction, which is sensitive to charge-noise-induced voltage fluctuations, and enable a measurable shift in the resonator signal under dispersive coupling, essentially providing a continuous classical readout of the exchange.\par
We demonstrated that, under realistic device parameters, the measurement time for achieving signal-to-noise ratio of unity can be on the order of 10 $\mu$s when using a high-quality resonator and a quantum-limited amplifier. Even in the absence of such an amplifier, sub-ms measurement times may be achievable with appropriate device engineering.\par
Importantly, we show that the valley degree of freedom, acting as a “spectator” state on the same physical qubit, produces negligible disturbance to the spin evolution. This enables the ST qubit to remain operational while the probe monitors the exchange. The measurement times are fast enough to allow tracking of low-frequency charge noise during quantum operations, enabling correction of slow drifts in qubit parameters during qubit operation. We outline some rudimentary ideas for a protocol in which detected shifts in the exchange interaction can be compensated in real time via gate voltage feedback and/or feedforward circuit updates to maintain qubit coherence. Future work will develop these correction protocols further.\par
By enabling fast, non-destructive, and spatially colocated charge noise measurement, this valley-mediated probe provides a new and potentially powerful tool for closed-loop stabilization of exchange-based qubits in silicon.

\acknowledgments

The authors acknowledge support from the Army Research Office (ARO) under Grant Number W911NF-23-1-0115. 

\appendix

\section{SNR Derivation}\label{app:SNR}
To derive the SNR for the probe, we first define the signal,
\begin{equation}
V(t)=\frac{\sqrt{R \hbar \omega_r }}{2}\, \delta j_\mrm{det}(t),
\end{equation}
where $\delta j_\mrm{det}(t)=2\sqrt{\kappa}\l(\re[\braket{a(t)}]-\re[\alpha_\mrm{ref}]\r)$ represents the deterministic and noiseless component of the signal, measured from a previously characterized reference signal value $\re[\alpha_\mrm{ref}]$ following Eq.~\eqref{eq:imrealpha}. 
The noisy signal is given by $V_N(t)=V(t)+\delta V(t)$, where $\delta V(t)$ is the stochastic noise present in the signal. 
Note that, as long as the system is stable, we have $\delta j_\mrm{det}(t)=0$; after a charge-noise event modifies the dispersive shift $\chi$, there will be a non-zero signal $\delta j_\mrm{det}(t)\neq 0$.
This signal is measured over a period of time, $t_m$, leading to the integrated signal, $I_N(t_m)$, defined as
\seq{
I_N(t_m) &= \int_0^{t_m} V_N(t) dt \\
&= \int_0^{t_m} V(t) dt + \int_0^{t_m} \delta V(t) d t \\
&= I(t_m) + \delta I(t_m),
}
where $I(t),\delta I(t)$ are defined analogously from $V(t),\delta V(t)$. By averaging the noisy signal over a measurement time $t_m$, the noisy contribution averages to zero as $t_m\rightarrow\infty$, i.e., $\delta I(t_m)/t_m \rightarrow 0$, and thus $I_N(t)/t_m$ will converge to the stationary value of the deterministic signal, namely $I_N(t_m)/t_m \rightarrow V(t_m)$. \par
The SNR is defined here by
\seq{
    \mathrm{SNR}^2(t) &= \frac{ \left| \mathbb{E}[I_N(t)] \right|^2}{\mathbb{E}[\delta I^2(t)]}. 
}
where we will assume that the noise $\delta V(t)$ is zero-mean, i.e.,  $\mathbb{E}[I_N(t)]=I(t)$. 
Assuming for simplicity that the signal is stabilized, we can write a small shift in the homodyne current as arising from a change in the dispersive shift, as shown in Eq.~\eqref{eq:dj_det_approx}, where in the steady-state we have 
\seq{
I^2 = \frac{\hbar \omega_r R \, t_m^2 \delta \chi^2 \braket{n}\,\kappa}{\kappa^2/4+\chi_0^2}.
}

Next, assuming stationarity of the noise, namely $C(t,t')=\mathbb{E}[\delta V(t)\delta V(t')]=C(t-t',0)$, we define the power spectral density (PSD) of the voltage noise as $S(f;t_m)= \int_0^{t_m} C(t,0) e^{-i2\pi f t} dt$.
In this work we are concerned with shot noise and Johnson noise, both of which have white noise power spectral densities, $S(f)=S_0$, over its bandwidth $\Delta f$. This results in a noise contribution obtained from the noise PSD, through
\seq{
\label{eq:deltaIPSD1}
\mathbb{E}[\delta I^2(t_m)] &= \int_0^{t_m}\int_0^{t_m} \mathbb{E}[\delta V(t) \delta V(t')] dt dt' \\
&= t_m^2 \int_0^{\Delta f} S(f) \, \text{sinc}^2 \left( \pi f t_m \right) df \\
&= \frac{\mrm{Si}(2\pi)}{\pi} S_0 t_m
}
where $\Delta f=1/t_m$ is the bandwidth of the noise, where $\mrm{Si}(x)=\int_0^x \text{sin}(x)/x \hspace{2pt}dx$ is the sine integral function, with $\mrm{Si}(2\pi)/\pi\approx 20/9$ .

When shot noise dominates the measurement noise, we have $\delta V(t)=\sqrt{R\hbar \omega_r}\xi(t)/2$. Then, $\mathbb{E}[\delta V(t) \delta V(t')]=R\hbar \omega_r \mathbb{E}[\xi(t)\xi(t')]/4=R\hbar \omega_r\delta(t-t')/4$ implies $\mathbb{E}[\delta I^2(t_m)]=R\hbar \omega_r t_m / 4$.
From here, we obtain the shot noise SNR
\seq{
\mrm{SNR}_s^2(t_m) &= \frac{4\,t_m \braket{n} \delta\chi^2 \kappa }{\kappa^2/4+\chi_0^2},\\
}
where imposing $\mrm{SNR}_s(t_m^\mrm{shot})=1$ yields Eq.~\eqref{eq:tm_vs_n}, namely $t_m^\mrm{shot} = (\kappa^2/4+\chi_0^2) / (4 \,\kappa\, \delta\chi_\mrm{min}^2 \braket{n})$.

Lastly, when thermal noise arising from the cryogenic amplifier dominates the measurement noise, we have  $S_0=4 k_B T R$~\cite{Johnson1928}, and using Eq.~\eqref{eq:deltaIPSD1} we obtain the SNR for thermal noise
\seq{
\mrm{SNR}_t^2(t_m) &= \frac{5}{9} \frac{\hbar \omega_r }{k_B T} \frac{t_m \delta \chi^2 \braket{n} \kappa}{\kappa^2/4+\chi_0^2}.
}
Setting $\mrm{SNR}_t(t_m^\mrm{cryo})=1$, we find
\seq{
t_m^\mrm{cryo} &= \frac{9}{5} \frac{k_B T}{\hbar \omega_r} \frac{\kappa^2/4+\chi_0^2}{\kappa \,\delta\chi^2 \braket{n}}\\
&\approx \frac{7 k_B T}{\hbar \omega_r} \times t_m^\mrm{shot},
}
shown in the main text in Eq.~\eqref{eq:tm_therm}.

\bibliographystyle{apsrev4-1}
\bibliography{refs_clean}

\end{document}